EUROPEAN ORGANIZATION FOR NUCLEAR RESEARCH

18 October 1995
CERN-PPE/95-149

# TESTS OF CPT SYMMETRY AND QUANTUM MECHANICS WITH EXPERIMENTAL DATA FROM CPLEAR

CPLEAR Collaboration

R. Adler[2], A. Angelopoulos[1], A. Apostolakis[1], E. Aslanides[11], G. Backenstoss[2], C.P. Bee[9], O. Behnke [17], A. Benelli[9], V. Bertin[11], F. Blanc[7,13], P. Bloch[4], P. Carlson[15], M. Carroll[9], J. Carvalho[5], E. Cawley[9], S. Charalambous[16], G. Chardin[14], M.B. Chertok[3], A. Cody[9], M. Danielsson[15], M. Dejardin[14], J. Derre[14], A. Ealet[11], B. Eckart[2], C. Eleftheriadis[16], I. Evangelou[8], L. Faravel [7], P. Fassnacht[11], C. Felder[2], R. Ferreira-Marques[5], W. Fetscher[17], M. Fidecaro[4], A. Filipčič[10], D. Francis[3], J. Fry[9], E. Gabathuler[9], R. Gamet[9], D. Garreta[14], H.-J. Gerber[17], A. Go[3,15], C. Guyot[14], A. Haselden[9], P.J. Hayman[9], F. Henry-Couannier[11], R.W. Hollander[6], E. Hubert[11], K. Jon-And[15], P.-R. Kettle[13], C. Kochowski[14], P. Kokkas[2], R. Kreuger[6], R. Le Gac[11], F. Leimgruber[2], A. Liolios[16], E. Machado[5], I. Mandić[10], N. Manthos[8], G. Marel[14], M. Mikuž[10], J. Miller[3], F. Montanet[11], T. Nakada[13], B. Pagels[17], I. Papadopoulos[16], P. Pavlopoulos[2], J. Pinto da Cunha[5], A. Policarpo[5], G. Polivka[2], R. Rickenbach[2], B.L. Roberts[3], T. Ruf[4], L. Sakeliou[1], P. Sanders[9], C. Santoni[2], M. Schäfer[17], L.A. Schaller[7], T. Schietinger[2], A. Schopper[4], P. Schune[14], A. Soares[14], L. Tauscher[2], C. Thibault[12], F. Touchard[11], C. Touramanis[4], F. Triantis[8], E. Van Beveren[5], C.W.E. Van Eijk[6], G. Varner[3], S. Vlachos[2], P. Weber[17], O. Wigger[13], M. Wolter[17], C. Yeche[14], D. Zavrtanik[10], D. Zimmerman[3]

and

J. Ellis[18], J.L. Lopez[19], N.E. Mavromatos[20], D. V. Nanopoulos[21].

**Abstract**

We use fits to recent published CPLEAR data on neutral kaon decays to $\pi^+\pi^-$ and $\pi e\nu$ to constrain the CPT–violation parameters appearing in a formulation of the neutral kaon system as an open quantum-mechanical system. The obtained upper limits of the CPT–violation parameters are approaching the range suggested by certain ideas concerning quantum gravity.

*(submitted to Phys. Lett. B)*

# 1 Introduction

The neutral kaon system provides one of the most sensitive laboratories for testing discrete symmetries and quantum mechanics at the microscopic level [1]. It is the only place where an experimental violation of CP symmetry has yet been seen, and may be used to probe the validity of CPT symmetry. This is a property of quantum field theory which follows from locality, causality and Lorentz invariance [2]. Violation of CPT would require us to revise one or more of these fundamental principles and it is therefore particularly important to search for it. Observations of the neutral kaon system have been used to test CPT symmetry, in particular within the context of conventional quantum mechanics via upper limits on a possible $K^0$–$\overline{K}^0$ mass difference [3, 4, 5].

Some approaches to quantum gravity suggest, according to Hawking [6], that quantum field theory should be modified in such a way that pure quantum-mechanical states evolve into mixed states, which would necessarily entail a violation of CPT [7]. This possibility may be analyzed using the formulation of open quantum-mechanical systems coupled to an unobserved environment [8]. In this framework, the observed system is described by a density matrix $\rho$ that obeys a modified quantum Liouville equation [8]

$$\dot{\rho} = i[\rho, H] + \not{\delta} H \rho \tag{1}$$

where $H$ is the usual quantum-mechanical Hamiltonian. The extra term $\propto \not{\delta} H$ would induce a loss of quantum coherence in the observed system, and hence a violation of CPT[8, 9, 10]. Since it is conjectured to arise from quantum-gravitational effects, the magnitude of $\not{\delta} H$ may be at most $\mathcal{O}(\mathrm{m_K}^2/\mathrm{M_{Pl}} \approx 2{\times}10^{-20}\ \mathrm{GeV})$, where $\mathrm{M_{Pl}} = 1.2{\times}10^{19}$ GeV is the gravitational mass scale obtained from Newton's constant: $\mathrm{M_{Pl}} = \mathrm{G_N}^{-\frac{1}{2}}$. An equation of the form (1) is supported by one interpretation of string theory [11], but could have more general applicability.

[1] University of Athens, Greece
[2] University of Basle, Switzerland
[3] Boston University, USA
[4] CERN, Geneva, Switzerland
[5] LIP and University of Coimbra, Portugal
[6] Technical University of Delft, Netherlands
[7] University of Fribourg, Switzerland
[8] University of Ioannina, Greece
[9] University of Liverpool, UK
[10] J. Stefan Inst. and Dep. of Physics, University of Ljubljana, Slovenia
[11] CPPM, IN2P3-CNRS et Université d'Aix-Marseille II, Marseille, France
[12] CSNSM, IN2P3-CNRS, Orsay, France
[13] Paul-Scherrer-Institut, Villigen, Switzerland
[14] CEA, DSM/DAPNIA, CE Saclay, France
[15] Royal Institute of Technology Stockholm, Sweden
[16] University of Thessaloniki, Greece
[17] ETH-IPP Zürich, Switzerland.
[18] CERN Theory Division, Geneva, Switzerland
[19] Department of Physics, Bonner Nuclear Lab., Rice University, 6100 Main Street, Houston, TX 77005, USA, supported in part by DOE grant DE-FG05-93-ER-40717
[20] P.P.A.R.C. Advanced Fellow, Dept. of Physics (Theoretical Physics), University of Oxford, 1 Keble Road, Oxford OX1 3NP, U.K.
[21] Center for Theoretical Physics, Department of Physics, Texas A&M University, College Station, TX 77843–4242, USA, and Astroparticle Physics Group, Houston Advanced Research Center (HARC), The Mitchell Campus, The Woodlands, TX 77381, USA, supported in part by DOE grant DE-FG05-91-ER-40633

In the case of the neutral kaon system, if the conservation of energy and strangeness are assumed, the open-system equation (1) introduces [8] three CPT-violation parameters $\alpha$, $\beta$ and $\gamma$, which are distinct from the $K^0$–$\overline{K}^0$ mass and decay width differences $\delta m = m_{\overline{K}^0} - m_{K^0}$, $\delta\Gamma = \Gamma_{\overline{K}^0} - \Gamma_{K^0}$. These parameters characterize the openness of an apparently isolated kaon to environmental effects which may be associated with quantum fluctuations in the space–time background. If present, they would cause the kaon density matrix to become mixed, via the suppression of off–diagonal entries in the $K^0$–$\overline{K}^0$ basis, thus modifying the rates of $K_S$ and $K_L$ decays as well as interference patterns. In order to maintain the hermiticity of the density matrix $\rho$ and the reality of the entropy, these parameters must obey the conditions [8]:

$$\alpha,\ \gamma > 0 \text{ and } \alpha\gamma > \beta^2 \tag{2}$$

Since these parameters would alter the time dependence of the neutral kaon state, they are best determined by studying the time evolution of neutral kaon decays, as measured by CPLEAR [12, 13]. Measurements [3, 14] at long lifetimes where the loss of quantum coherence is increased can be used as additional constraints. The sensitivity reached is in the range of interest to the above-mentioned quantum-gravitational ideas.

## 2 The CPLEAR measurement of $K^0$ and $\overline{K}^0$ decay rate asymmetries

The CPLEAR experiment [15] is designed to study CP and T violation in the neutral kaon system by measuring time dependent decay rate asymmetries of CP and T conjugate processes. Initially pure $K^0$ and $\overline{K}^0$ states are produced concurrently via the antiproton annihilations $(p\overline{p})_{\rm rest} \rightarrow K^0K^-\pi^+$ and $(p\overline{p})_{\rm rest} \rightarrow \overline{K}^0K^+\pi^-$. The strangeness of the neutral kaon is tagged by the charge sign of the accompanying kaon. The experimental set–up enables us to measure the neutral kaon decay rates in the region of 0 to 20 $\tau_S$. However, the main statistical significance of our measurements is obtained in the region below 10 $\tau_S$. The correlated analysis of the decays to $\pi^+\pi^-$ and $\pi e\nu$ is crucial for the determination of the parameters $\alpha$, $\beta$, $\gamma$ as explained below.

The analysis of the $\pi^+\pi^-$ and $\pi e\nu$ final states are described in [12, 13]. The measured decay rates into the $\pi^+\pi^-$ final state are shown in figure 1 separately for $K^0$ and $\overline{K}^0$, demonstrating visually the known CP violation effect. The decay rate difference is measured using the time dependent asymmetry

$$A_{2\pi}(\tau) = \frac{N_{\overline{K}^0\rightarrow\pi^+\pi^-}(\tau) - a \times N_{K^0\rightarrow\pi^+\pi^-}(\tau)}{N_{\overline{K}^0\rightarrow\pi^+\pi^-}(\tau) + a \times N_{K^0\rightarrow\pi^+\pi^-}(\tau)} \tag{3}$$

In this asymmetry all acceptances common to $K^0$ and $\overline{K}^0$ cancel, reducing systematic uncertainties. The normalization factor $a$ of the $K^0$, $\overline{K}^0$ rates is fitted simultaneously with the CPT violation parameters $\alpha$, $\beta$, $\gamma$ following the procedure described in [12]. The measured rates $N_{\overline{K}^0\rightarrow\pi^+\pi^-}(\tau)$ and $N_{K^0\rightarrow\pi^+\pi^-}(\tau)$ have been corrected for regeneration [12].

For evaluating the $A_{2\pi}$ asymmetry (see Eq. 5 below) a precise knowledge of $\Delta m$ and the $K_S$–$K_L$ decay width difference $\Delta\Gamma$ is needed. These two parameters are normally calculated assuming the time evolution of states according to conventional quantum mechanics. The decay width difference is dominated by the $K_S$ mean life, which is measured at early decay times because of statistical reasons, and is therefore not much affected by the modifications of the open quantum-mechanical formalism which accumulate with time. This is not the case for the mass difference $\Delta m$ and results of experiments using conventional quantum mechanical formalism cannot easily be used. However, by using the data of the CPLEAR experiment, the value of $\Delta m$ can be obtained in the framework of the open quantum-mechanical formalism, by analysing $A_{\Delta m}$ [13]:

$$A_{\Delta m}(\tau) = \frac{[N_{\overline{K^0}\to e^-\pi^+\overline{\nu}}(\tau) + N_{K^0\to e^+\pi^-\nu}(\tau)] - [N_{\overline{K^0}\to e^+\pi^-\nu}(\tau) + N_{K^0\to e^-\pi^+\overline{\nu}}(\tau)]}{[N_{\overline{K^0}\to e^-\pi^+\overline{\nu}}(\tau) + N_{K^0\to e^+\pi^-\nu}(\tau)] + [N_{\overline{K^0}\to e^+\pi^-\nu}(\tau) + N_{K^0\to e^-\pi^+\overline{\nu}}(\tau)]} \qquad (4)$$

Because of the special form of $A_{\Delta m}$, the systematic errors are reduced and all terms linear in the regeneration amplitude cancel [13].

## 3 Description of the Fit

We use the formalism for the time-dependent $K^0$ and $\overline{K^0}$ decay rates to two pions and semileptonic final states presented in [16], where the open-system evolution equation (1) is solved perturbatively in the small parameters $\alpha$, $\beta$, $\gamma$ and $\varepsilon$. Direct CP violation in kaon decay amplitudes at the level currently allowed by experiment does not affect our analysis and is ignored in the formalism. Since $A_{2\pi}(\tau)$ contains an important term $\propto |\varepsilon|^2$, the following full second-order formula for $A_{2\pi}(\tau)$ is required:

$$\begin{aligned} A_{2\pi}(\tau) &= \Big\{ 2|\varepsilon|\cos\phi + 4\widehat{\beta}\sin\phi\cos\phi - 8\widehat{\alpha}\sin\phi\cos\phi(|\varepsilon|\sin\phi - 2\widehat{\beta}\cos^2\phi) \\ &\quad -2\sqrt{|\varepsilon|^2 + 4\widehat{\beta}^2\cos^2\phi}\; e^{\frac{1}{2}(1/\tau_S - 1/\tau_L)\tau} \left[\cos(\Delta m\tau - \phi - \delta\phi) + \frac{2\widehat{\alpha}}{\tan\phi}X_\alpha\right]\Big\} \\ &\quad / \left\{1 + e^{(1/\tau_S - 1/\tau_L)\tau}\left[\widehat{\gamma} + |\varepsilon|^2 - 4\widehat{\beta}^2\cos^2\phi - 4\widehat{\beta}|\varepsilon|\sin\phi\right]\right\} \end{aligned} \qquad (5)$$

where $\widehat{\alpha}$, $\widehat{\beta}$, $\widehat{\gamma}$ are scaled variables $\alpha/\Delta\Gamma$, $\beta/\Delta\Gamma$, $\gamma/\Delta\Gamma$ and the CP impurity parameter $\varepsilon$ is given with the proper phase convention by :

$$\varepsilon = |\varepsilon|e^{-i\phi} = \frac{\mathrm{Im M_{12}}}{\frac{1}{2}|\Delta\Gamma| + \mathrm{i}\Delta\mathrm{m}} \quad \text{with} \quad \phi \equiv \mathrm{atan}\frac{2\Delta\mathrm{m}}{\Delta\Gamma}$$

Furthermore, to the desired order we may write

$$\begin{aligned} \delta\phi &\equiv \mathrm{atan}\left(-\frac{2\widehat{\beta}\cos(\phi)}{|\varepsilon|}\right) \\ X_\alpha &\equiv \cos\delta\phi\sin(\Delta m\tau - \phi) - \frac{1}{2}|\Delta\Gamma|\tau\tan\phi\cos(\Delta m\tau - \phi - \delta\phi) \\ &\quad + \sin\phi\cos(\Delta m\tau - 2\phi - \delta\phi) \end{aligned}$$

On the other hand, the following expansion of $A_{\Delta m}(\tau)$ to first order, where we assumed the validity of the $\Delta S = \Delta Q$ rule, is sufficient for our purposes:

$$A_{\Delta m}(\tau) = \frac{2e^{-\frac{1}{2}(1/\tau_S + 1/\tau_L)\tau}\left[\cos\Delta m\tau + \frac{2\widehat{\alpha}}{\tan\phi}(\sin\Delta m\tau - \Delta m\tau\cos\Delta m\tau)\right]}{e^{-\tau/\tau_L}(1 + 2\widehat{\gamma}) + e^{-\tau/\tau_S}(1 - 2\widehat{\gamma})} \qquad (6)$$

We see from equations (5) and (6) that the open quantum mechanical formalism induces a time dependence which can be distinguished from that of conventional quantum mechanics, even with a CPT–violating $K^0$–$\overline{K^0}$ mass difference [5] incorporated. Fitting simultaneously the two asymmetries, $A_{2\pi}$ (Eq. 5) and $A_{\Delta m}$ (Eq. 6) to the CPLEAR data (Fig. 2 and Fig. 3), we obtain:

$$\begin{aligned} \alpha &= (-0.3 \pm 3.3) \times 10^{-17}\,\mathrm{GeV} \\ \beta &= (2.0 \pm 2.5) \times 10^{-19}\,\mathrm{GeV} \\ \gamma &= (-0.5 \pm 1.4) \times 10^{-18}\,\mathrm{GeV} \end{aligned}$$

$$|\varepsilon| \;=\; (2.34 \pm 0.08) \times 10^{-3} \quad \text{and} \quad \Delta\text{m} = (526.3 \pm 4.2) \times 10^{7} \hbar \text{s}^{-1} \tag{7}$$

Since the effects of the open quantum-mechanical formalism accumulate with time, the best limits on the CPT violation parameters $\alpha$, $\beta$, $\gamma$ are obtained by using in addition to the CPLEAR asymmetries, other measurements of $\text{K}_\text{L}$ decays which refer essentially to long decay times [14, 17]. In the formalism used here the parameters $|\eta_{+-}|$ and $\delta_L$ as defined in [3] are given for long decay times by:

$$|\eta_{+-}|^2 = \widehat{\gamma} + |\varepsilon|^2 \frac{\cos(\phi - 2\delta\phi)}{\cos\phi \cos^2\delta\phi} \tag{8}$$

$$\delta_L = 2|\varepsilon| \cdot \cos\phi - 4\widehat{\beta}\cos\phi \sin\phi \tag{9}$$

Equations (8) and (9) are used as additional constraints, with the values $|\eta_{+-}| = (2.30 \pm 0.035) \times 10^{-3}$ from [14] and $\delta_L = (3.27 \pm 0.12) \times 10^{-3}$ (average calculated by the Particle Data Group [3]). Fitting simultaneously the two CPLEAR asymmetries with the two constraints (Eqs. 8 and 9), we obtain:

$$\begin{aligned} \alpha &= (-0.5 \pm 2.8) \times 10^{-17}\,\text{GeV} \\ \beta &= (2.5 \pm 2.3) \times 10^{-19}\,\text{GeV} \\ \gamma &= (1.1 \pm 2.5) \times 10^{-21}\,\text{GeV} \end{aligned}$$

$$|\varepsilon| = (2.32 \pm 0.06) \times 10^{-3} \quad \text{and} \quad \Delta m = (526.3 \pm 3.5) \times 10^{7} \hbar \text{s}^{-1} \tag{10}$$

The main constraint on the parameter $\gamma$ is given by the comparison of the measurement of the amplitude of the interference pattern (Eq. 5), which defines the value of $|\varepsilon|$, with the measurement of $|\eta_{+-}|^2$ (Eq. 8), where $\widehat{\gamma}$ enters linearly. The result of the fit is shown in Fig. 2 and Fig. 3 as a solid line.

Systematic errors are derived by varying the resolution, the background, and regeneration corrections within their uncertainties[12, 13] and are added in quadrature to the statistical errors. The possibility that the observed CP violation is mimicked by the discussed CPT violation effects ($\varepsilon = 0$) is excluded by the measurements (see also [9, 10, 16]).

## 4 Final Results

The likelihood distribution of the three parameters $\alpha$, $\beta$ and $\gamma$ is approximated by a gaussian which takes also into account their correlation, as given by the correlation matrix of the fit (see table 1). The 90% confidence level upper limits on each of the parameters are obtained by implementing the positivity constraints (Eq. 2) and integrating over the likelihood distribution of the other parameters:

$$\alpha < 4.0 \times 10^{-17}\,\mathrm{GeV} \qquad |\beta| < 2.3 \times 10^{-19}\,\mathrm{GeV} \qquad \gamma < 3.7 \times 10^{-21}\,\mathrm{GeV}\ . \tag{11}$$

| | $\alpha$ | $\beta$ | $\gamma$ | $\lvert\varepsilon\rvert$ | $\Delta m$ |
|---|---|---|---|---|---|
| $\alpha$ | 1.0 | +0.2 | –0.4 | +0.7 | +0.6 |
| $\beta$ | | 1.0 | +0.4 | +0.2 | –0.3 |
| $\gamma$ | | | 1.0 | –0.7 | –0.5 |
| $\lvert\varepsilon\rvert$ | | | | 1.0 | +0.4 |
| $\Delta m$ | | | | | 1.0 |

Table 1: The correlation coefficients for the parameters $\alpha$, $\beta$, $\gamma$, $|\varepsilon|$ and $\Delta m$ of the fit

We note that these limits approach the range $\mathcal{O}(\mathrm{m_K}^2/\mathrm{M_{Pl}} \approx 2\times 10^{-20}\,\mathrm{GeV})$, which is of interest for testing ideas based on quantum gravity.

**Acknowledgments**

This work was supported by the following institutions: the French CNRS/Institut National de Physique Nucléaire et de Physique des Particules (IN2P3), the French Commissariat à l'Energie Atomique (CEA), the Greek General Secretariat of Research and Technology, the Netherlands Foundation for Fundamental Research on Matter (FOM), the Portuguese JNICT and INIC, the Ministry of Science and Technology of the Republic of Slovenia, the Swedish Natural Science Research Council, the Swiss National Science Foundation, the UK Particle Physics and Astronomy Research Council (PPARC), the US National Science Foundation (NSF) and the US Department of Energy (DOE).

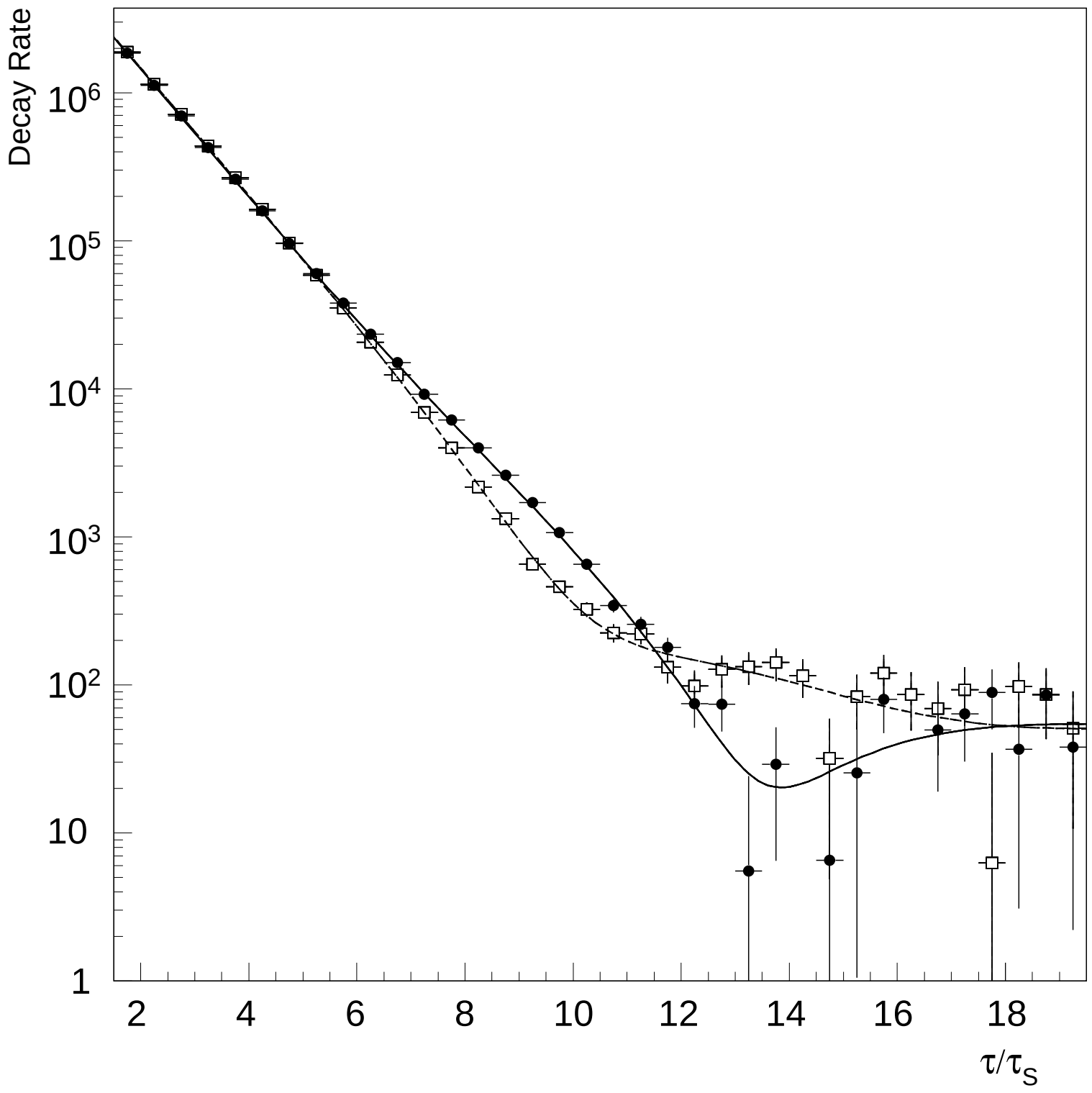

Figure 1: Acceptance-corrected decay rate of $K^0$(□) and $\overline{K}^0$(●) into $\pi^+\pi^-$. The lines are the expected rates (Eq. 1) when the PDG–94 values are used for $\Delta m$, $\tau_S$, $\tau_L$ and $\eta_{+-}$.

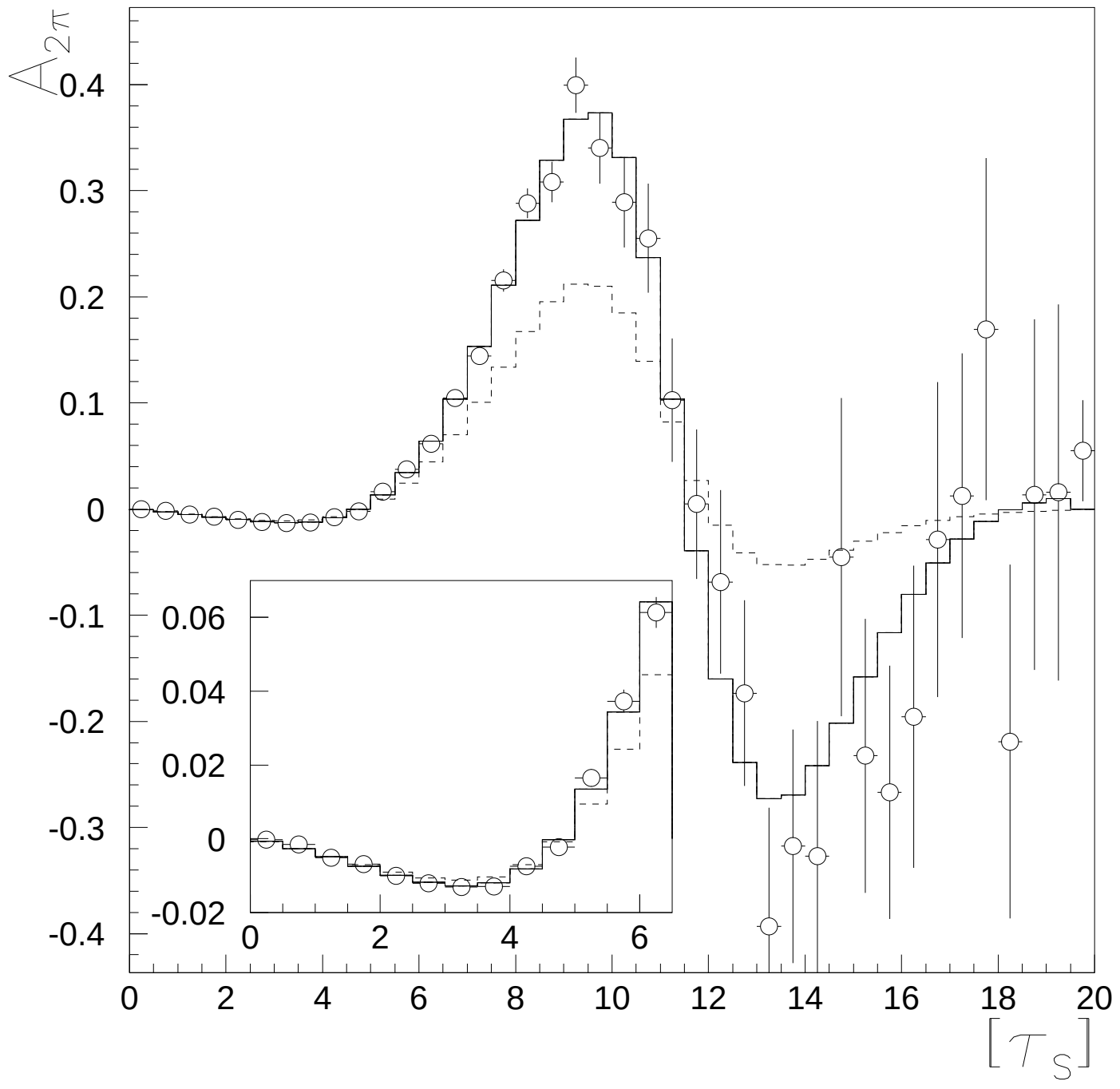

Figure 2: Asymmetry $A_{2\pi}$ as a function of the decay eigentime. The dots represent the CPLEAR data on $\pi^+\pi^-$ decays. The inset displays the data at short decay times. The solid line is the result of our fit. The dashed line represents $A_{2\pi}$ with positive values of $\alpha$, $\beta$, $\gamma$ which are for illustration 10 times larger than the actual limits obtained (Eq. 11). The shape of the curve is insensitive to the value of the parameter $\gamma$.

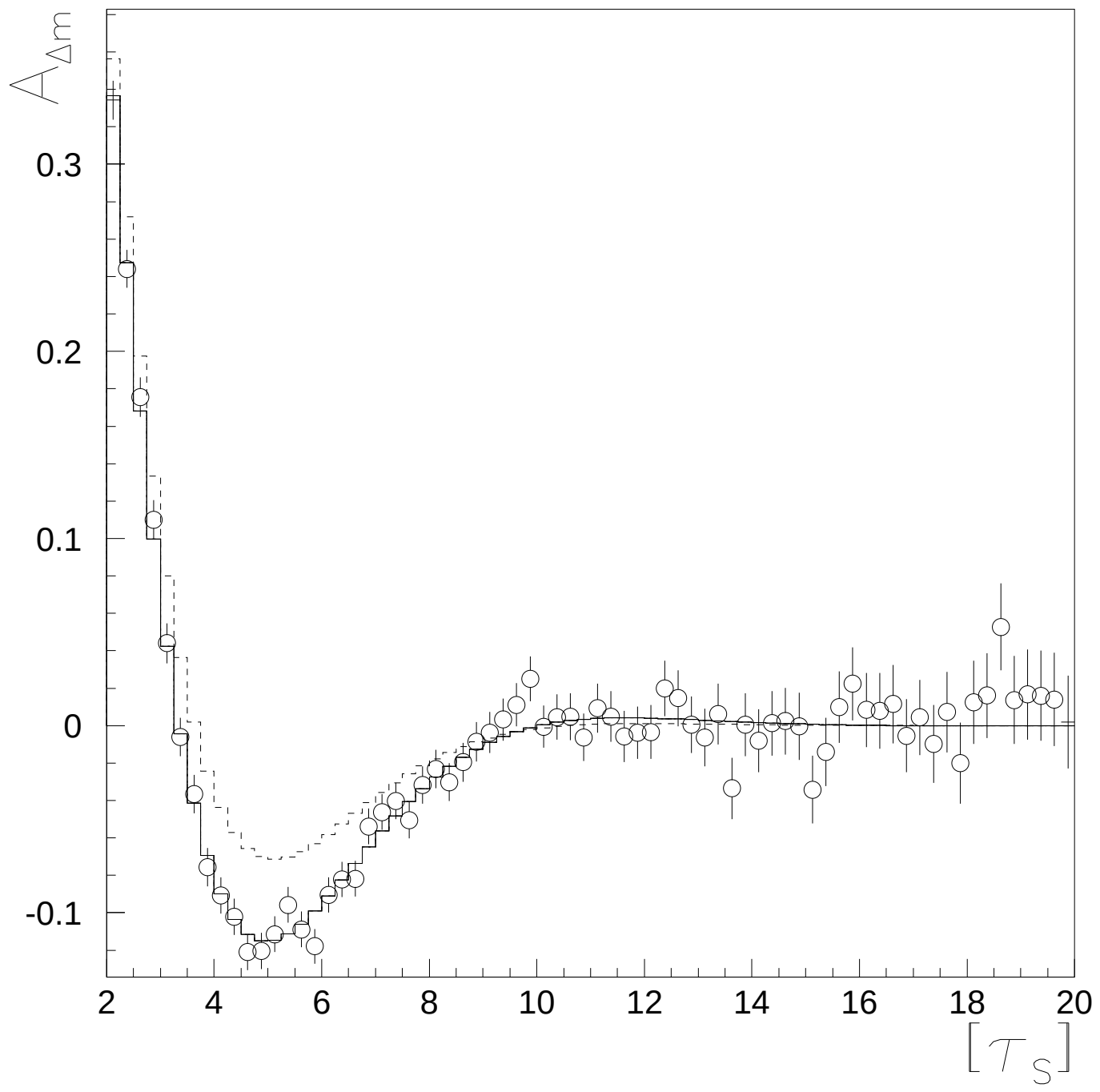

Figure 3: Asymmetry $A_{\Delta m}$ showing the CPLEAR data (dots) on $\pi e \nu$ decays. The solid line is the result of our fit. The dashed line represents $A_{\Delta m}$ with positive values of $\alpha$, $\beta$, $\gamma$ which are for illustration 10 times larger than the actual limits obtained (Eq. 11). The shape of the curve depends mainly on the parameter $\alpha$, apart from $\Delta m$.